\let\SpringerBibliographyStyle\bibliographystyle
\renewcommand{\bibliographystyle}[1]{\SpringerBibliographyStyle{sn-basic-qmi}}
\documentclass[pdflatex,sn-basic]{sn-jnl}
\let\bibliographystyle\SpringerBibliographyStyle
\usepackage[T1]{fontenc}
\usepackage{lmodern}
\usepackage{graphicx}
\usepackage{amsmath,amssymb,amsfonts}
\usepackage{amsthm}
\usepackage[title]{appendix}
\usepackage{booktabs}
\usepackage{braket}
\usepackage{upgreek}
\usepackage{xspace}
\usepackage{microtype}
\usepackage{etoolbox}
\usepackage{xcolor}

\let\cite\citep

\makeatletter
\DeclareRobustCommand\onedot{\futurelet\@let@token\@onedot}
\def\@onedot{\ifx\@let@token.\else.\null\fi\xspace}
\def\eg{\emph{e.g}\onedot}
\def\etal{\emph{et al}\onedot}
\makeatother

\newcommand{\fref}[1]{Fig.~\ref{#1}}
\newcommand{\sref}[1]{Sec.~\ref{#1}}
\newcommand{\V}[1]{\ensuremath{\mathbf{#1}}}
\newcommand{\vg}{\V{g}}
\newcommand{\vp}{\V{p}}
\newcommand{\vx}{\V{x}}

\theoremstyle{thmstyleone}
\newtheorem{theorem}{Theorem}

\let\springerabstract\abstract
\makeatletter
\RenewDocumentEnvironment{abstract}{+b}{%
  \springerabstract{#1}%
  \global\let\@abstract\@abstract
}{}
\makeatother

\let\springerfigure\figure
\let\endspringerfigure\endfigure
\RenewDocumentEnvironment{figure}{O{tbp}}{%
  \ifstrempty{#1}{\springerfigure[tbp]}{\springerfigure[#1]}%
}{\endspringerfigure}

\hypersetup{bookmarksdepth=3, hidelinks}
\begin{document}

\title{Representation Geometry and Inductive Biases in Quantum Classifiers across Classical Visual Benchmarks}

\author*[1]{\fnm{Yu} \sur{Yamaoka}}\email{yamaoka.yuu.qiqb@osaka-u.ac.jp}
\author[2]{\fnm{Weng Ian} \sur{Chan}}
\author[3]{\fnm{Kosuke} \sur{Mitarai}}

\affil*[1]{\orgdiv{Center for Quantum Information and Quantum Biology}, \orgname{The University of Osaka}, \orgaddress{\street{1-2 Machikaneyama}, \city{Toyonaka}, \postcode{560-0043}, \state{Osaka}, \country{Japan}}}

\affil[2]{\orgdiv{SANKEN}, \orgname{The University of Osaka}, \orgaddress{\street{1-5 Yamadaoka}, \city{Suita}, \postcode{565-0871}, \state{Osaka}, \country{Japan}}}

\affil[3]{\orgdiv{Graduate School of Engineering Science}, \orgname{The University of Osaka}, \orgaddress{\street{1-3 Machikaneyama}, \city{Toyonaka}, \postcode{560-8531}, \state{Osaka}, \country{Japan}}}

\begin{abstract}
Quantum Machine Learning (QML) has a theoretical advantage over classical algorithms for specific synthetic data that is computationally hard.
However, it remains unclear whether such theoretical advantages play a meaningful role for realistic classical data, such as natural images.
Quantum structural properties induced by unitary operations or entanglement introduce inductive biases that differ from those of classical models, potentially leading to distinct learning behavior on real-world data.
In a fully quantum learning pipeline, classical real-world data are first encoded into quantum states via a quantum feature mapping, analogous to a classical feature extractor, and a quantum classifier then classifies the resulting states.
Quantum feature mappings generate representations that are difficult to realize classically, while also inducing representation-level biases that do not necessarily match the structure of realistic data.
To enable a fair comparison at the classifier level, we focus on the classifier and use classical feature extractors that already provide powerful representations for natural images.
By controlling the data scale and supervision strength, we analyze how the unitary structure of quantum classifiers affects generalization, and show that their effectiveness depends strongly on the geometry of the input representations.
These results indicate that quantum classifiers benefit from classical features that are aligned with quantum feature spaces.
Our analysis shows that kernel alignment identify representations compatible with quantum feature spaces, providing insights into translating classical models into quantum algorithms.
\end{abstract}


\maketitle

\section{Introduction}
\label{sec:intro}

Recent progress in machine learning relies on large models, high-resolution inputs, and large-scale datasets, which increases computational cost.
This trend raises fundamental questions about the long-term sustainability of relying solely on classical digital hardware.
To address this challenge, researchers begin exploring alternative computational paradigms.
Quantum computing in machine learning is one potential approach.
Unlike classical computation, which processes deterministic bit transitions, quantum computation manipulates superposed states and evolves them through unitary transformations, effectively operating on exponentially large state spaces in parallel.
It thus provides a fundamentally different computational substrate for large-scale learning.
Prior work on quantum machine learning~(QML) demonstrates computational advantages of quantum models over classical learners on carefully constructed learning tasks, such as those derived from the discrete logarithm problem~ \cite{liu2021rigorous}.
In addition to such hardness-based advantages, QML studies have also highlighted distinctive inductive properties of quantum models, including enhanced expressivity~\cite{Abbas2021,havlivcek2019supervised} and implicit regularization induced by unitary constraints~\cite{jiang2023strong,caro2022generalization}.

However, whether these theoretical advantages translate into practical improvements on natural image data remains an open question.
Applying QML to such data is nontrivial, as classical inputs must be encoded into quantum states and learning procedures must be implemented using unitary quantum operations. These requirements impose structural constraints that differ fundamentally from those of classical architectures.
As circuit depth increases, these constraints can also lead to well-known optimization problems, including barren plateaus~\cite{larocca2025barren}.
Therefore, evaluating QML on realistic image data requires understanding what inductive biases quantum classifiers introduce and how these biases affect predictive performance and trainability. In this work, we systematically characterize the classifier-level inductive biases in vision classification across diverse datasets~\cite{netzer2011reading,qmnist-2019,medmnistv2,fashion,stl,CIFAR}.

The inductive biases of quantum models depend on where quantum components are introduced in the learning pipeline: in the feature mapping stage, the classifier stage, or both~\cite{havlivcek2019supervised,Abbas2021,cong2019quantum}.
While fully quantum models are an attractive long-term objective, a fair evaluation of quantum inductive bias requires controlled input conditions at the classifier stage and assessment on realistic data rather than synthetic datasets constructed for a quantum model.
When both the feature extractor and classifier are quantum, the feature mapping may produce representations that are especially well matched to the quantum classifier, making it difficult to isolate classifier-level inductive bias because the observed behavior depends on both components.
For natural images, strong self-supervised classical backbones already provide powerful visual representations~\cite{DINO-origin,oquab2023dinov2,simeoni2025dinov3,he2016deep,radford2021learning}, while the advantage of quantum feature mappings over such backbones remains unclear.
Therefore, we use the classical feature extractor and replace only the classifier to isolate the quantum model's contribution for a fair evaluation on classical image tasks.

Our study tests two hypotheses.
First, the unitary constraint in quantum circuits may serve as an implicit regularizer by enforcing transformations that preserve the norm of quantum states, thereby improving generalization under limited supervision.
Second, the geometry of quantum state spaces may introduce inductive biases that appear more clearly in small-data or weakly supervised settings.
We carefully control model capacity and circuit depth to ensure a fair and meaningful evaluation.
In particular, we design quantum classifiers with parameter counts comparable to those of their classical counterparts while avoiding excessively deep circuits that are prone to optimization problems, such as barren plateaus.
This controlled setup allows us to attribute observed differences to inductive bias rather than to mismatched model capacity or unstable training dynamics.
By isolating the quantum classifier from feature-representation effects, we identify when quantum models behave similarly to classical models and when unitary constraints help or hinder performance.

Our contributions are as follows:

\begin{itemize}

\item \textbf{Theoretical perspective on quantum inductive bias.}
We formalize conditions under which quantum classifiers can operate comparably to classical models on natural image representations, and analyze how unitary constraints act as implicit regularization under limited supervision.

\item \textbf{Controlled empirical evaluation across supervision regimes.}
We systematically evaluate quantum classification heads across multiple classical feature extractors, dataset sizes, and weakly supervised settings. This allows us to disentangle effects arising from quantum expressivity and those induced by unitary structural constraints.

\item \textbf{Fair comparison under matched capacity and depth control.}
We match parameter counts between quantum and classical classifiers and restrict circuit depth to mitigate optimization pathologies such as barren plateaus, ensuring that observed differences reflect inductive bias rather than model capacity or instability.

\end{itemize}

In modern vision systems dominated by strong self-supervised backbones, clarifying whether alternative computational paradigms provide complementary inductive biases is a first step toward practical quantum vision systems.

\section{Related work}
\label{sec:related}

\textbf{Quantum feature map and kernel}
Quantum feature maps are among the most widely studied approaches in quantum machine learning.
Havlíček \etal~\cite{havlivcek2019supervised} proposed embedding classical data into high-dimensional quantum states and performing classification via kernel methods.
This idea was further formalized by Schuld and Killoran~\cite{schuld2019quantum}, who showed that many quantum models can be interpreted as implicit kernel methods.
More recently, Liu \etal~\cite{liu2021rigorous} established a provable quantum advantage for a specific learning problem using quantum kernel estimation.

\textbf{Variational quantum classifier}
Another major line of research focuses on variational quantum classifiers (VQCs),
where parametrized quantum circuits are trained using classical optimizers.
Early work by Farhi and Neven~\cite{farhi2018classification}
demonstrated classification using shallow quantum circuits.
Subsequent studies investigated the expressive power of variational quantum models,
showing that quantum neural networks can represent complex decision boundaries in high-dimensional Hilbert spaces~\cite{Abbas2021}.

\textbf{Expressivity and trainability theory}
Theoretical studies have analyzed the representational capacity and trainability of quantum models.
Huang \etal~\cite{Huang2021} established information-theoretic bounds on quantum advantage in learning.
Caro \etal~\cite{caro2022generalization} investigated the generalization behavior of quantum models trained on small datasets.
At the same time, several works have highlighted optimization challenges in variational quantum circuits,
such as the barren plateau phenomenon, where gradients vanish exponentially with system size~\cite{cerezo2022challenges,larocca2025barren}.

\textbf{Application to real-world vision data}
Applications of quantum machine learning to computer vision tasks have also been explored.
Cong \etal~\cite{cong2019quantum} introduced quantum convolutional neural networks,
while Henderson \etal~\cite{henderson2020quanvolutional} proposed quanvolutional layers for image feature extraction.
Hybrid classical–quantum pipelines combining classical feature extractors with quantum models have also been studied in transfer learning settings~\cite{mari2020transfer}.

Despite these efforts, the empirical benefits of quantum components on natural image datasets remain unclear.
Most existing studies rely on small-scale datasets or synthetic benchmarks,
and it remains an open question whether quantum models provide practical advantages
when combined with strong classical feature representations.

\vspace{-3mm}
\section{Method}
\label{sec:method}

\subsection{Quantum circuit learning}

\paragraph{Quantum state in Bra-Ket Notation}
In this paper, we use Dirac's bra-ket notation to represent a quantum state, and the appendix explains this notation in detail.
Any one-qubit pure quantum state is
\(
\ket{\psi}= \alpha\ket{0}+\beta\ket{1} \
(\alpha,\beta\in\mathbb{C},\ |\alpha|^2+|\beta|^2=1)
\)
based on
\(
\ket{0}=\begin{bmatrix}1\\0\end{bmatrix},\
\ket{1}=\begin{bmatrix}0\\1\end{bmatrix}
\) as a computational basis.
The zero state of the $n$-qubit is
\(
\ket{0}^{\otimes n}=\ket{0}\otimes\cdots\otimes\ket{0}
\).
The tensor product $\otimes$ represents the direct product of multi-qubit states.

\paragraph{Unitary time evolution in a quantum circuit}
A quantum circuit is a computational model based only on unitary matrices.
We obtain the quantum state $\ket{\psi}=U_L\cdots U_2U_1\ket{0}^{\otimes n}$ though gates $U_1U_2\cdots$ from zero state $\ket{0}^{\otimes n}$.

\begin{figure}[t]
\vspace{-3mm}
    \includegraphics[width=\linewidth]{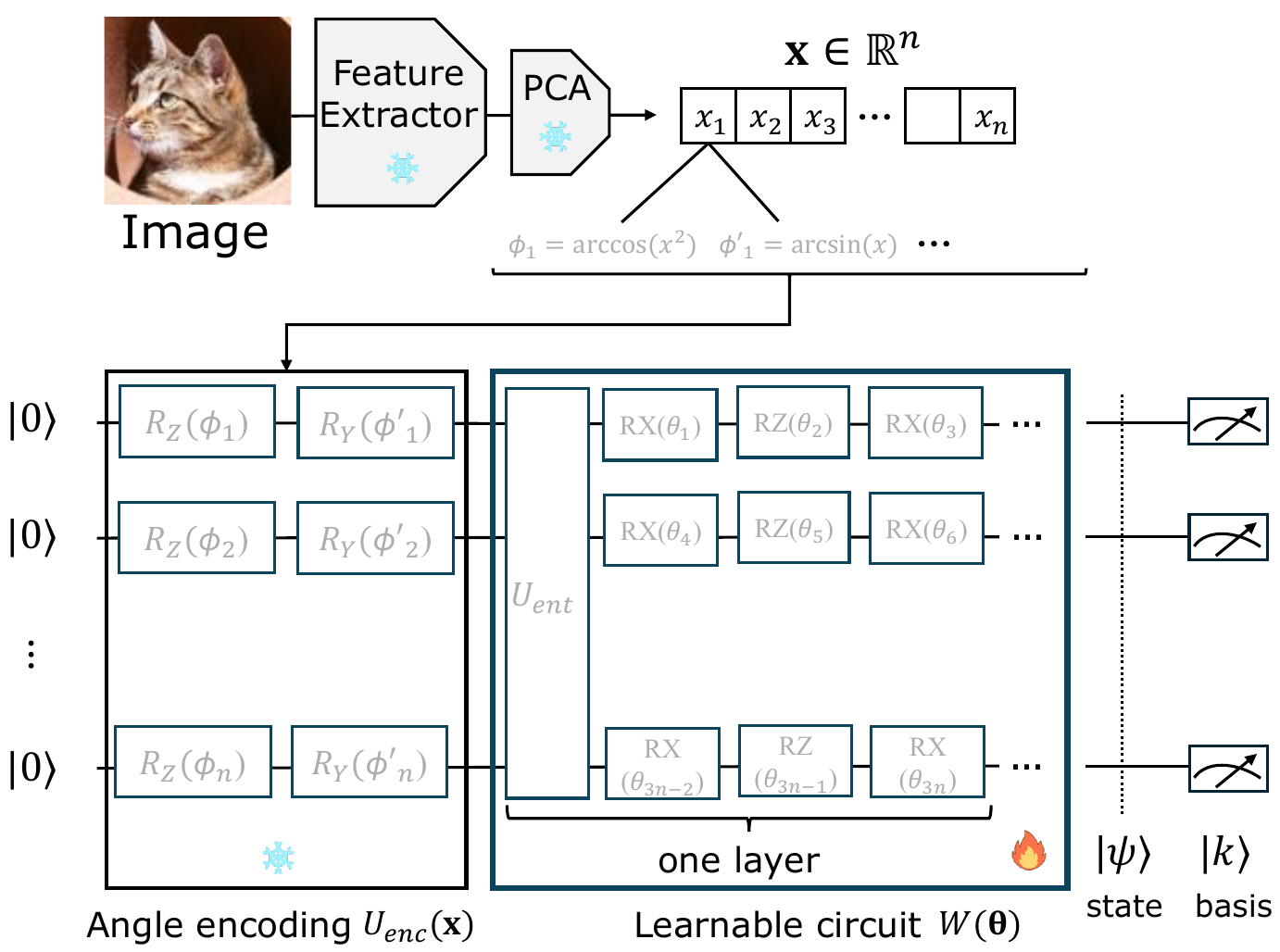}
    \caption{Quantum classification pipeline using classical features. $U_{ent}$ is a fixed entangling unitary applied across the qubits.}
    \label{fig:method_quantum_circuit}
\vspace{-3mm}
\end{figure}

\paragraph{Encoding and learning}
As shown in Fig.~\ref{fig:method_quantum_circuit}, quantum circuit encodes classical features $\mathbf{x}\in\mathbb{R}^n$ obtained thorugh a classical feature extractor~\cite{radford2021learning,he2016deep,DINO-origin,oquab2023dinov2,simeoni2025dinov3,howard2017mobilenets} and PCA~\cite{PCApedregosa2011scikit} to quantum circuit by angle encoding~\cite{havlivcek2019supervised}.
We preserve the normalization of the amplitudes from the input data as a function of the angles $\phi$ and $\phi'$.
This naturally introduces inverse trigonometric functions; the initial angle for the $i$-th qubit is $  \theta_i \;=\; \arcsin\bigl(x_{i}\bigr),  \phi_i \;=\; \arccos\bigl(x_{i}^{\,2}\bigr)$.

The trainable unitary matrix $W(\boldsymbol{\uptheta})$ is appended after the feature encoding $U(\vx)\ket{0}$. The circuit $W(\boldsymbol{\uptheta})$ is capable of adapting through training, regardless of the initial state $\ket{0}$.
\[
  U(\vx,\boldsymbol{\uptheta}) \;=\; W(\boldsymbol{\uptheta})\,U_{\mathrm{enc}}(\vx).
\]

The resulting parameterized quantum state $\ket{\psi(\vx,\boldsymbol{\uptheta})}$ is $U(\vx,\boldsymbol{\uptheta})\,\ket{0}^{\otimes n}$.

In the optimization step, we compute the gradient with respect to the circuit parameters.
We describe the gradient computation in detail in the appendix.

\vspace{0mm}
\paragraph{Generalization based on unitary}
Because unitary operators satisfy $U^\dagger U = I$, they preserve the norm of the quantum state.
This property leads to the generalization bound stated below.

\begin{theorem}[generalization vs.\ circuit depth]
For an $n$-qubit quantum circuit consisting of $T$ unitary operations, the generalization performance of the converged parameters $\boldsymbol{\hat\uptheta}$ after training has upper-bounded in terms of the dataset size $N$ as follows~\cite{caro2022generalization}:

\begin{equation}
\label{eq:gene}
\mathrm{generalization}(\boldsymbol{\hat\uptheta})\
\;\le\;
\mathcal{O}\left(\sqrt{\frac{T \log T}{N}}\right),
\end{equation}
\end{theorem}

For the same task, a quantum model use $T=\mathrm{poly}(n)$ trainable gates,
whereas a classical model requires $O(2^{n})$ parameters in the worst case to achieve comparable expressivity results in \sref{subsec: generalization}.

\vspace{0mm}
\subsection{Kernel geometry induced by the quantum classifier}

In our setting, the image representation $\mathbf{x}$ is fixed by a classical
feature extractor. A parameterized quantum circuit then maps the classical feature vector
$\mathbf{x}\in\mathbb{R}^n$ to a quantum state $\ket{\psi(\mathbf{x},\boldsymbol{\theta})}=U(\mathbf{x},\boldsymbol{\theta})\ket{0}^{\otimes n}$.

This mapping implicitly defines a feature representation in the Hilbert space.
We characterize the representation through the quantum kernel.

\[
K_Q(\mathbf{x}_i,\mathbf{x}_j)
=
|\langle\psi(\mathbf{x}_i,\boldsymbol{\theta})|\psi(\mathbf{x}_j,\boldsymbol{\theta})\rangle|^2 .
\]

This kernel corresponds to the inner product between quantum feature states.
It measures the similarity between samples after the quantum feature transformation.

\vspace{0mm}
\paragraph{Kernel Target Alignment}
We evaluate the quality of the induced feature representation using
Kernel Target Alignment (KTA).
Given a kernel matrix $K$ and a one-hot label matrix
$Y\in\{0,1\}^{N\times C}$, we define the label kernel as
$T=YY^\top$ and the centering matrix as $H=I_N-\frac{1}{N}\mathbf{1}\mathbf{1}^\top$.
Here, $N=\min(1024, N_{\mathrm{train}})$ denotes the number of training samples used to construct the
kernel matrices and $C$ is class number.
KTA is defined as

\[
\mathrm{KTA}(K,Y)
=
\frac{\langle HKH,HTH\rangle_F}
{\|HKH\|_F\,\|HTH\|_F}.
\]

KTA measures how well the feature-induced similarity structure
aligns with the class labels.
Higher KTA indicates that samples from the same class become more
similar in the feature space.

\vspace{-3mm}
\paragraph{Quantum vs. classical feature geometry}

We compare the representation induced by the quantum circuit with
that of the original classical features.
Let $K_V$ denote the kernel computed from the classical feature
vectors and $K_Q$ denote the kernel induced by the quantum states.
We analyze the change in alignment.

\[
\Delta\mathrm{KTA} = \mathrm{KTA}_Q - \mathrm{KTA}_V .
\]

A positive $\Delta\mathrm{KTA}$ indicates that the quantum-induced similarity structure is more aligned with the class labels than the original classical feature geometry.
In Sec.~\ref{sec:result}, we study how this change relates to
classification accuracy across datasets and feature extractors results in \sref{subsec: Cross-Dataset} and \sref{subsec: Parameter efficiency of quantum}.

\section{Experiment and Result}
\label{sec:result}

\subsection{Generalization by Unitary constraints}
\label{subsec: generalization}

\begin{figure}[t]
\vspace{0mm}
    \includegraphics[width=\linewidth]{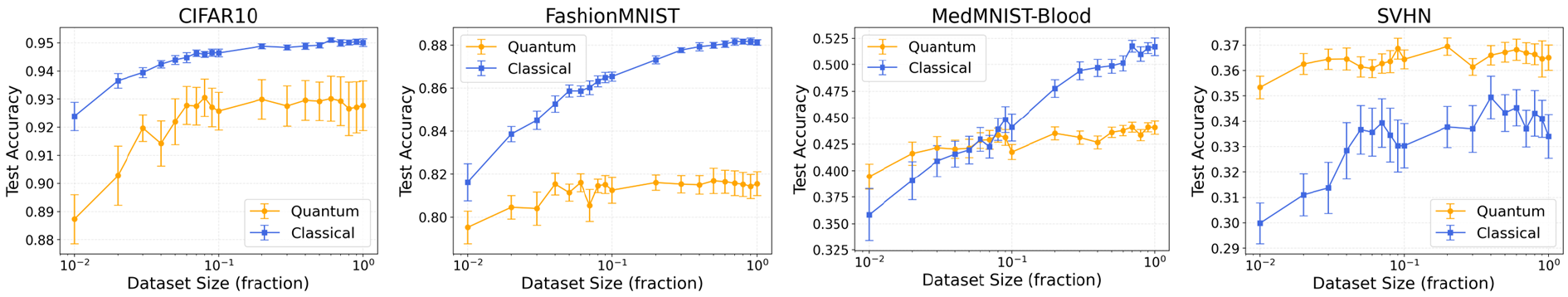}
    \caption{Mean test accuracy with $95$\% confidence interval as error bars across different dataset sizes (higher is better).}
    \label{fig:test_accuracy}
\vspace{-3mm}

\vspace{0mm}
    \includegraphics[width=\linewidth]{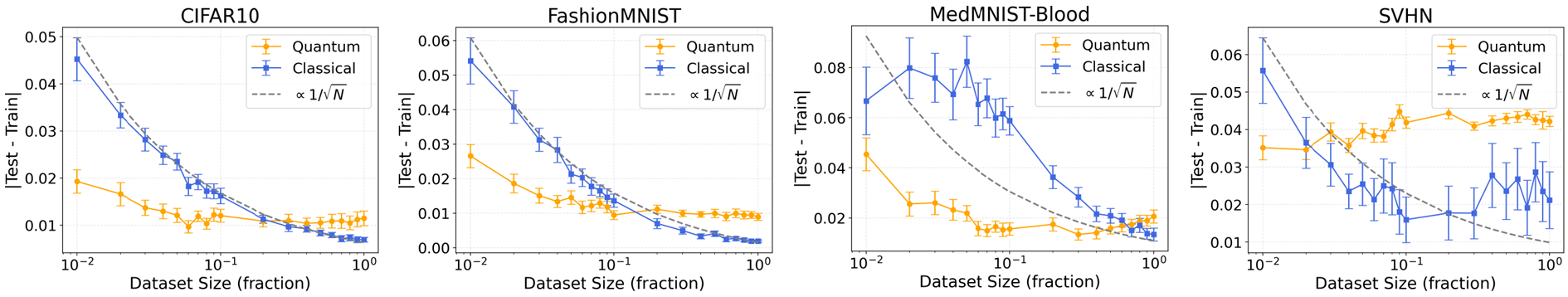}
    \caption{Mean generalization with $95$\% confidence interval as error bars across different dataset sizes (lower is better). Generalization is defined as the difference between training and test accuracies.}
    \label{fig:test_gene}
\vspace{-3mm}
\end{figure}

\noindent{\textbf{Objective.}}
This experiment evaluates our first hypothesis introduced in Sec.~\ref{sec:intro}:
The unitary structure of quantum classifiers may act as an implicit regularizer and improve generalization under limited supervision.
To test it, we adopt learning from label proportions (LLP), a weakly supervised setting in which the model observes only class proportions over groups of samples rather than individual instance labels.
This setting reduces the amount of supervision while preserving the underlying image classification task, allowing us to examine whether quantum classifiers are more robust than classical classifiers when label information is limited. We define our LLP loss in the appendix.

\noindent{\textbf{Experimental setting.}}
We use the first four classes in each dataset due to the high computational cost of the quantum simulation.
We set the bag size to ten, which defines the weakly supervised setting of learning from label proportions.
We use qulacs-gpu (v0.6.11)\footnote{Qulacs is a Python/C++ library for fast simulation of parametric quantum circuits~\cite{suzuki2021qulacs}.~<\url{https://github.com/qulacs/qulacs}>, last accessed on September 4, 2025.} as a quantum simulator and DINO v2\footnote{Self-Supervised Vision Transformers with DINO~\cite{oquab2023dinov2}.~<\url{https://github.com/facebookresearch/dinov2}>, last accessed on September 4, 2025.} ViT-S/14 model as the feature extractor.
The classical classifier consists of three fully connected layers with the ReLU activation function~\cite {ReLU}.
The number of learnable parameters is hidden layer size $h$, input size $n$, class number $C$, as $(n+1)h+(h+1)h+(h+1)C$ from the hyperparameters.
The number of parameters of the quantum circuit is $3 \times n \times depth$, calculated from the depth of the three parametric gates RX–RZ–RX and the qubit number $n$.
Under the condition of class number $C=4$, in order to make the number of parameters to be updated in quantum and classical close, we set $n=4, h=15, depth=32$, obtaining: the number of learnable parameters $\boldsymbol{\hat\uptheta}$ is $379$ for classical and $384$ for quantum.
We use 95\% confidence intervals computed over fifty independently trained models initialized with different random seeds to account for variability arising from parameter initialization.
\\

\noindent{\textbf{Result.}}
Fig.~\ref{fig:test_accuracy} shows the test accuracy of the training dataset fraction across four benchmarks (CIFAR10~\cite{CIFAR}, FashionMNIST~\cite{fashion}, MedMNIST-Blood~\cite{medmnistv2}, and SVHN~\cite{netzer2011reading}).
Fig.~\ref{fig:test_gene} reports the corresponding generalization gap, defined as $|\mathrm{TestAcc} - \mathrm{TrainAcc}|$, together with a reference scaling proportional to $1/\sqrt{N}$ following Eq.~\ref{eq:gene}.

Across all datasets, increasing the dataset fraction improves test accuracy for both the classical and quantum classifiers.
However, their relative behavior differs significantly across data fractions.
At low fractions, the classical classifier often achieves slightly higher raw test accuracy on CIFAR10 and FashionMNIST.
On the other hand, the quantum classifier is marginally better on MedMNIST-Blood and SVHN, suggesting that the two approaches are overall competitive across datasets.
Nevertheless, the quantum classifier consistently exhibits a smaller generalization gap (Fig.~\ref{fig:test_gene}), indicating reduced overfitting.
This result suggests that the unitary constraint acts as an implicit regularizer.

As the dataset size $N$ increases, the generalization gap of the classical model decreases approximately, following the expected scaling $\mathcal{O}\!\left(\frac{1}{\sqrt{N}}\right)$ as indicated by the dashed reference line in Fig.~\ref{fig:test_gene}.
On CIFAR10 and FashionMNIST, the classical head ultimately achieves higher asymptotic accuracy, while the quantum plateaus earlier, suggesting a capacity limitation induced by the unitary constraint.
In contrast, on SVHN, quantum consistently outperforms the classical classifier across dataset fractions (Fig.~\ref{fig:test_accuracy}), and exhibits lower variance and more stable generalization behavior.

While the generalization gap decreases with $N$ for both models, the quantum classifier starts from a noticeably smaller gap, suggesting a smaller prefactor in the expected $\mathcal{O}(1/\sqrt{N})$ scaling.
One possible explanation is that, for certain tasks, a quantum model can learn with $T=\mathrm{poly}(n)$ trainable gates. In contrast, a classical model can need $\mathcal{O}(2^{n})$ parameters in the worst case to match the same level of expressivity.

\subsection{Cross-Dataset and Cross-Feature Comparison}
\label{subsec: Cross-Dataset}

\noindent{\textbf{Objective.}}
This experiment evaluates our second hypothesis: the geometry induced by quantum state spaces interacts with the input representation, leading quantum classifiers to behave differently across datasets and feature extractors.
To examine this effect, we compare quantum and classical classifier heads under controlled training conditions while varying both the dataset and the feature extractor.

\noindent{\textbf{Experimental setting.}}
We use seven feature extractors and ten datasets, and keep hyperparameters and training procedures identical to ensure a controlled comparison.

We set $\mathrm{frac}=0.1$ for CIFAR-10, FashionMNIST, MedMNIST-Derma, QMNIST, and SVHN, considering the computational cost of quantum simulation.
We use $\mathrm{frac}=0.5$ for Food101 and STL-10 and $\mathrm{frac}=1.0$ for Flowers102, Oxford-IIIT Pet, and MedMNIST-Retina due to their smaller training-set sizes.
We keep these fractions across quantum and classical models, and evaluate all models on the full test set of the corresponding four-class subset.

We consider four-class subsets for each dataset due to the computational cost of quantum simulation.
The bag size is fixed at 1 in this setting, corresponding to fully supervised instance-level labels.
For quantum training, we run $1000$ optimization iterations with 10 number of random seeds.
We use seven-dimensional features and qubit $n=7$, which is the largest feature dimension we could practically evaluate given the computational cost of quantum simulation. We also match the parameter counts of the quantum and classical classifiers as closely as possible. Feature extractors include CLIP, DINO variants, MobileNetV3-Large, ResNet50, and PCA-based features.

\begin{figure}[]
    \includegraphics[width=\linewidth]{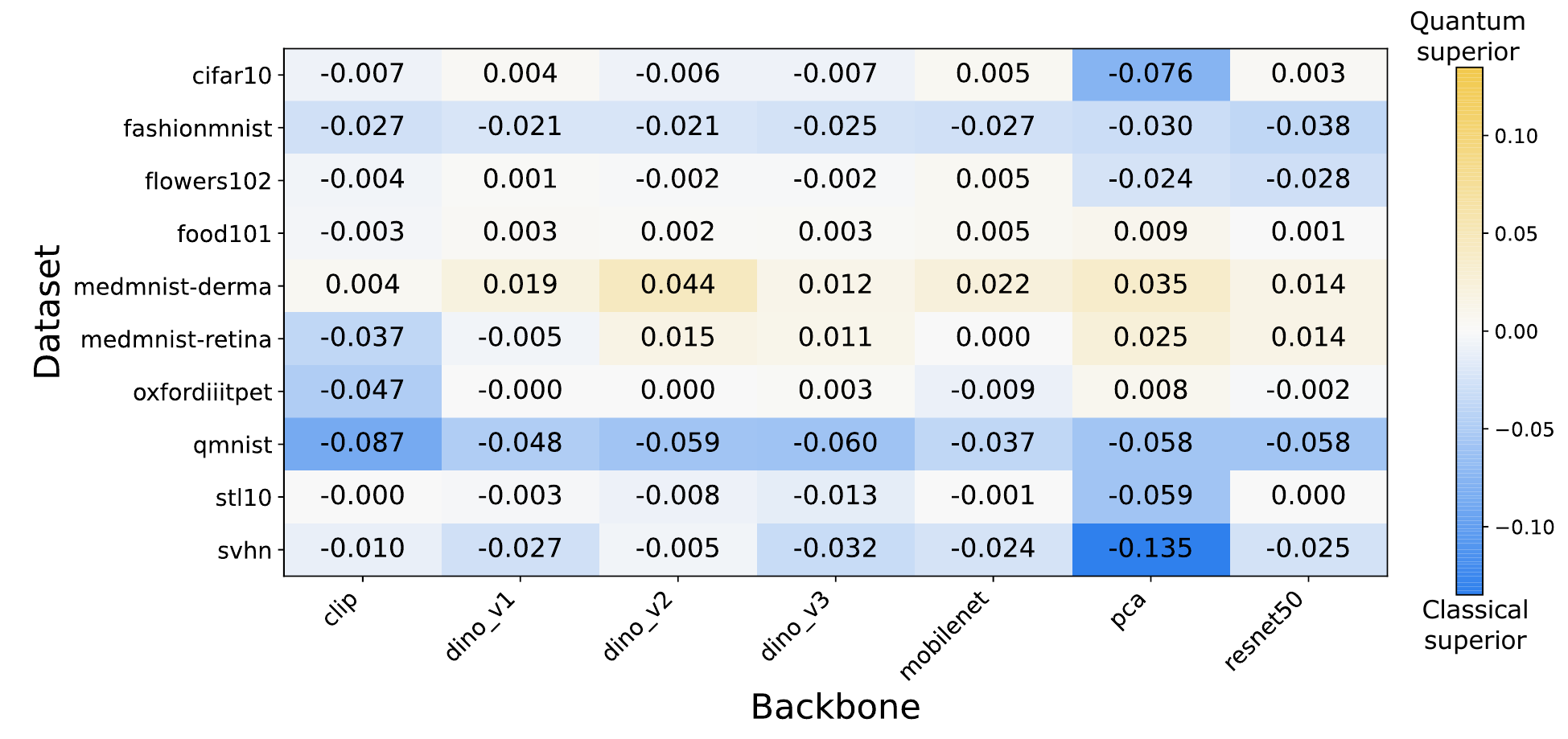}
    \caption{Test accuracy on cross-dataset and cross-feature comparison.}
    \label{fig:cross_acc}
\end{figure}

\noindent{\textbf{Result.}}
Figure~\ref{fig:cross_acc} shows the difference in test accuracy between the classical and quantum classifiers across datasets and feature extractors, where positive values indicate higher accuracy for the quantum classifier, whereas negative values indicate higher accuracy for the classical classifier.
Overall, the classical classifier achieves higher accuracy in most settings. The gap is particularly large on Flowers102, where several backbones (e.g., DINO v1 and ResNet50) show differences larger than 0.2. Moderate advantages for the classical model also appear on datasets such as FashionMNIST and QMNIST.

In contrast, the classical classifier achieves higher accuracy in most dataset–feature combinations. Quantum improvements occur only in a limited number of settings, including MedMNIST-Derma with DINOv2 and PCA features and MedMNIST-Retina with PCA features. The magnitude of these improvements remains modest, whereas several settings show substantially larger advantages for the classical classifier.
These results indicate that replacing the classical classifier with a unitary quantum classifier does not consistently improve performance on standard image features. The results also suggest that the effectiveness of the quantum model depends strongly on the dataset and the choice of feature representation.

\begin{figure}[t]
\vspace{0mm}
    \includegraphics[width=\linewidth]{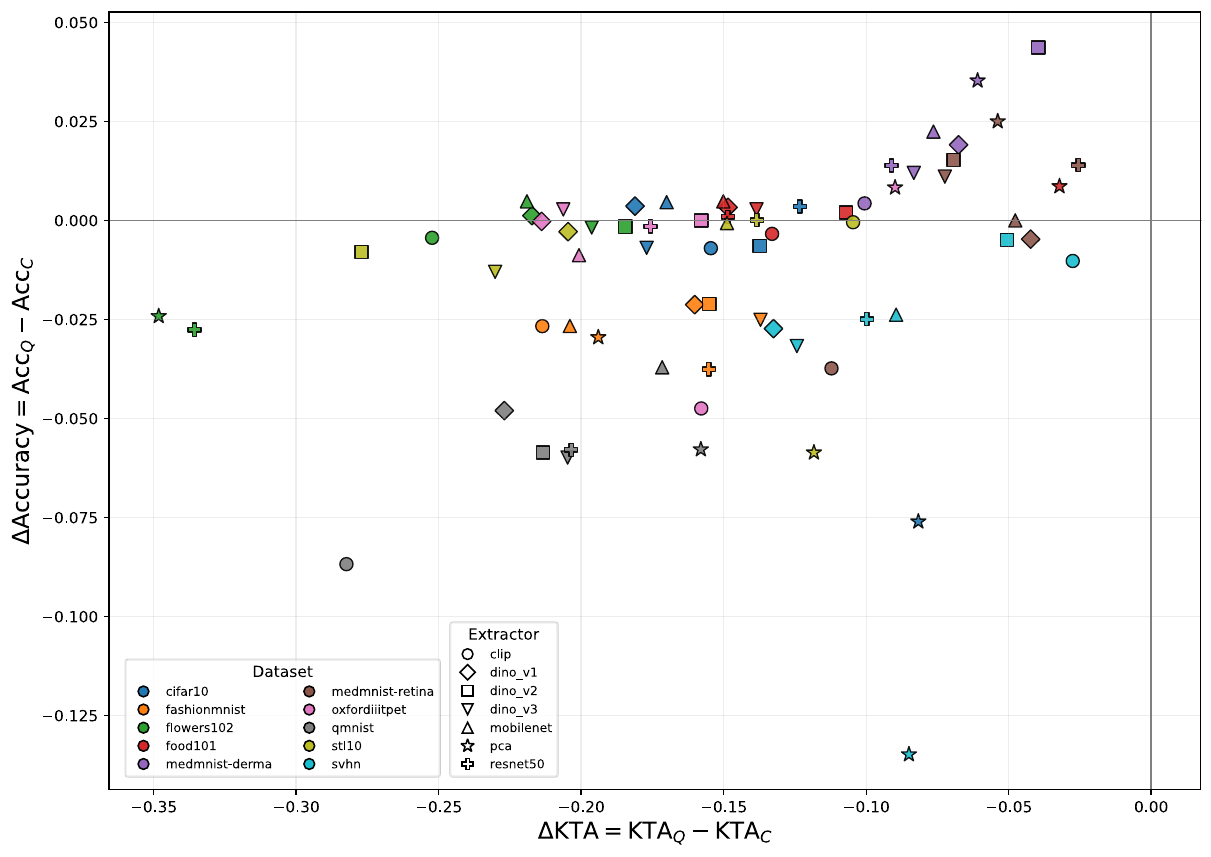}
    \caption{
    Relationship between the change in kernel-target alignment $\Delta\mathrm{KTA}$ and the test accuracy difference $\Delta\mathrm{Acc}$.
    Each point corresponds to a dataset--backbone pair, with colors indicating datasets and marker shapes indicating backbones.
    A positive correlation (Pearson \(r=0.312\), Spearman \(\rho=0.429\)) indicates that performance differences follow changes in representation alignment.
    }
    \label{fig:kta_vs_delta_acc}
\vspace{-3mm}
\end{figure}

\noindent{\textbf{Relationship between geometry change and performance.}}
Figure~\ref{fig:kta_vs_delta_acc} plots the relationship between the change in kernel-target alignment (KTA) and the change in test accuracy between quantum and classical classifiers across all datasets and feature extractors.
Each point corresponds to a dataset--backbone pair, and the axes represent
\(\Delta\mathrm{KTA} = \mathrm{KTA}_{Q} - \mathrm{KTA}_{V}\) and
\(\Delta\mathrm{Acc} = \mathrm{Acc}_{Q} - \mathrm{Acc}_{V}\), respectively.
We observe a clear positive correlation between these quantities
(Pearson \(r=0.312\), Spearman \(\rho=0.429\)).

This trend indicates that performance differences between quantum and classical classifiers are strongly related to the alignment between the induced representation geometry and the class labels.
When the quantum classifier increases KTA relative to the classical classifier, the test accuracy tends to improve as well.
Conversely, when the quantum representation reduces alignment with class labels, the performance gap becomes negative.

Importantly, the absolute KTA of the classical representation alone does not explain the performance gap (see Fig.~\ref{fig:kta_vs_delta_acc}).
Instead, the change in alignment induced by the classifier is more significant.
These results suggest that quantum classifiers do not simply improve weak feature geometries; rather, they reshape the feature space's similarity structure.
Performance gains arise when the induced quantum geometry better matches the data's class structure.

\vspace{0mm}
\subsection{Parameter efficiency of quantum}
\label{subsec: Parameter efficiency of quantum}

\begin{figure}[]
\vspace{-3mm}
\centering
    \includegraphics[width=0.94\linewidth]{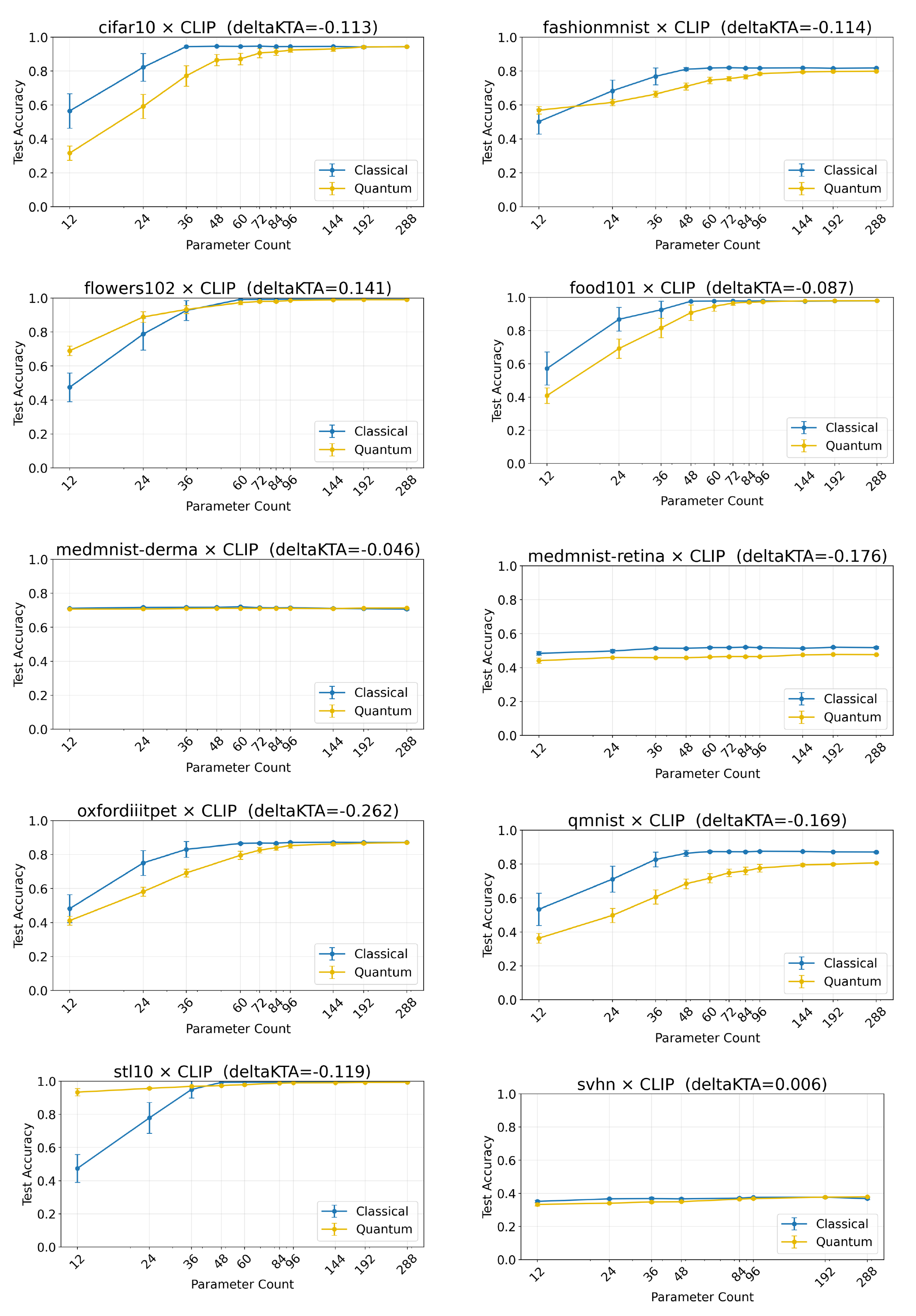}
    \caption{Test accuracy versus the number of trainable parameters for each dataset using CLIP-extracted image features and a quantum classifier head.}
    \label{fig:param}
\vspace{-3mm}
\end{figure}

We further study how test accuracy changes as the number of trainable parameters increases.
Figure~\ref{fig:param} shows the resulting accuracy curves.
In these cases, the quantum model reaches competitive accuracy with fewer parameters
than the classical model. This observation suggests that when the quantum feature
representation aligns well with the label structure (high $\Delta\mathrm{KTA}$),
the circuit can represent the task using a smaller number of trainable parameters.
However, such favorable cases are rare. Among a total of 70 dataset–feature combinations
we evaluated, only a small fraction exhibit this behavior.
Overall, these observations indicate that quantum circuits do not uniformly
improve performance, but they can provide parameter-efficient representations
for specific feature geometries.

\section{Limitation and Conclusion}
\label{sec:conlcusion}

\vspace{0mm}
\paragraph{Limitations.}
This work studies classifier-level quantum inductive bias in a controlled simulation setting.
Because classical simulation of quantum circuits is computationally expensive, our experiments are limited to four-class subsets, reduced training fractions, and small circuits with $n=4$ qubits at ~\fref{fig:test_accuracy} or $n=7$ at ~\fref{fig:cross_acc}.
The observed trends therefore characterize a low-dimensional regime and should be further validated with more qubits, more classes, and full datasets.

We also compress classical visual features, such as DINOv2 and CLIP embeddings, to four dimensions using PCA to match the quantum input size.
This enables a controlled comparison between quantum and classical classifier heads under identical input dimensionality, but it introduces an information bottleneck and does not represent the performance of full-dimensional classical probes.
Accordingly, our results should be interpreted as an analysis of quantum heads on compressed visual representations, rather than as evidence of general quantum advantage over modern classical vision models.

Finally, quantum classifiers are not universally superior in our experiments; classical MLPs outperform them in several settings.
This highlights the importance of analyzing when quantum circuit structure is beneficial, rather than assuming a universal advantage.
Future work should explore larger circuits, quantum-aware feature mappings, and evaluations beyond classical simulation.

\paragraph{Conclusion.}
Quantum computing is currently transitioning from the Noisy Intermediate-Scale Quantum (NISQ) era towards the development of fault-tolerant quantum computers (FTQC).
Quantum machine learning is expected to benefit from the high computational capacity of FTQC systems and may offer new opportunities to improve machine learning models.

In this work, we investigated the inductive biases of quantum classifiers for classical image features.
We constructed a competitive quantum classifier with a comparable number of parameters to classical baselines and evaluated it across diverse datasets and feature extractors.
Under controlled settings of data scale and supervision strength, the quantum model showed stable generalization behavior induced by unitary operators.
Across 70 evaluated datasets--feature combinations, the quantum classifier achieved modest accuracy or parameter-efficiency advantages in only a small subset, while the classical classifier performed better in most settings.
Furthermore, our analysis showed that kernel alignment provides a useful indicator for identifying representations compatible with quantum feature spaces.
Quantum classifiers showed no clear advantage on conventional image classification tasks without an underlying source of classical computational hardness when sufficient many trainable parameters and training data are available.
However, quantum classifiers show advantages when the classical feature geometry aligns with quantum feature spaces.
This observation offers a practical guideline for selecting classical feature representations when translating classical models into quantum algorithms.

\section{Acknowledgments}
Y.Y. was supported by JST COI-NEXT (Grant No. JPMJPF2014) and JSPS KAKENHI (Grant No. 25K24367).
W.C. was supported by JST (Grant Number JPMJFS2125).
K.M. was supported by JST FOREST (Grant No. JPMJFR232Z) and JSPS KAKENHI (Grant Nos. 23H03819 and 24K16980).

\section*{Author contributions statement}
Y.Y. conceived the experiments. Y.Y., K.M. constructed an inference and learning pipeline and conducted the experiments.
Y.Y., W.C. wrote the manuscript. All authors analyzed the results and reviewed the manuscript.

\section*{Data availability}
The datasets generated during the current study are available from the corresponding author on reasonable request.
Source codes can be found in this Github(\url{https://anonymous.4open.science/r/Quantum_Classifier-F176/}), and updates will be made periodically.

\bibliography{main}

\begin{appendices}
\renewcommand{\theHfigure}{appendix.\arabic{figure}}
\renewcommand{\theHtable}{appendix.\arabic{table}}
\clearpage

\renewcommand{\thetable}{\Alph{table}}
\renewcommand{\thefigure}{\Alph{figure}}
\setcounter{figure}{0}
\setcounter{table}{0}

\section{Appendix}

\subsection{Quantum circuit}
\paragraph{Quantum state in Bra-Ket Notation}
In this paper, we use Dirac's bra-ket notation for the representation of a quantum state.
Any one-qubit pure quantum state is
\(
\ket{\psi}= \alpha\ket{0}+\beta\ket{1}
(\alpha,\beta\in\mathbb{C},\ |\alpha|^2+|\beta|^2=1)
\)
based on
\(
\ket{0}=\begin{bmatrix}1\\0\end{bmatrix},\
\ket{1}=\begin{bmatrix}0\\1\end{bmatrix}
\) as a computational basis.

The zero state of the $n$-qubit is
\(
\ket{0}^{\otimes n}=\ket{0}\otimes\cdots\otimes\ket{0}
\).
The tensor product $\otimes$ represents the direct product of multi-qubit states.

\paragraph{Unitary time evolution in a quantum circuit}
A quantum circuit is a computational model based on only unitary matrices.
We get quantum state $\ket{\psi}=U_L\cdots U_2U_1\ket{0}^{\otimes n}$ though gates $U_1U_2\cdots$ from zero state $\ket{0}^{\otimes n}$

\paragraph{One-qubit gate}
The Pauli matrices
\(X=\begin{bmatrix}0&1\\1&0\end{bmatrix},\
Y=\begin{bmatrix}0&-j\\ j&0\end{bmatrix},\
Z=\begin{bmatrix}1&0\\0&-1\end{bmatrix}\) are the fundamental operations.
In this section, $j$ denotes the imaginary unit.
We define $R_X$, $R_Y$, and $R_Z$ gates as
\[
R_{A}(\theta)\;=\;\exp\!\left(-\frac{j\theta}{2}A\right)\quad (A\in\{X,Y,Z\}).
\]

We encode the features obtained from the backbone into a quantum circuit by the operation $R_Y(\theta), R_Z(\phi)$ as below.
\[
R_Y(\theta)\ket{0}=\cos\frac{\theta}{2}\ket{0}+\sin\frac{\theta}{2}\ket{1},\qquad
R_Z(\phi)\ket{0}=e^{-j\phi/2}\ket{0},\ \ R_Z(\phi)\ket{1}=e^{+j\phi/2}\ket{1}.
\]
A global complex phase factor of the entire state does not affect observable quantities.

\paragraph{Multi-qubits and entanglement}
In multi-qubits, we generate entanglement state $\ket{\psi}\neq \ket{\psi_1}\otimes\cdots\otimes\ket{\psi_n}$ by control gate~\eg $\mathrm{CNOT}$ gate.
Weight unitary with learning parameter $W(\boldsymbol{\uptheta})$ includes entanglement.

\paragraph{Measurement of quantum state}
We use a Pauli $Z$ basis for measurement, 0 or 1 probability of the $i$-th qubit is as below by using the expectation value $\langle Z^{(i)}\rangle=\bra{\psi}Z^{(i)}\ket{\psi}\in[-1,1]$

\[
\Pr(\text{$i$-th qubit}=1)=\frac{1-\langle Z^{(i)}\rangle}{2},\quad
\Pr(\text{$i$-th qubit}=0)=\frac{1+\langle Z^{(i)}\rangle}{2}
\]
We use the expectation value of the loss function of the classification model.

\subsubsection{Feature Encoding}
\label{sec.encode}

\begin{figure}[htb]
    \includegraphics[width=\linewidth]{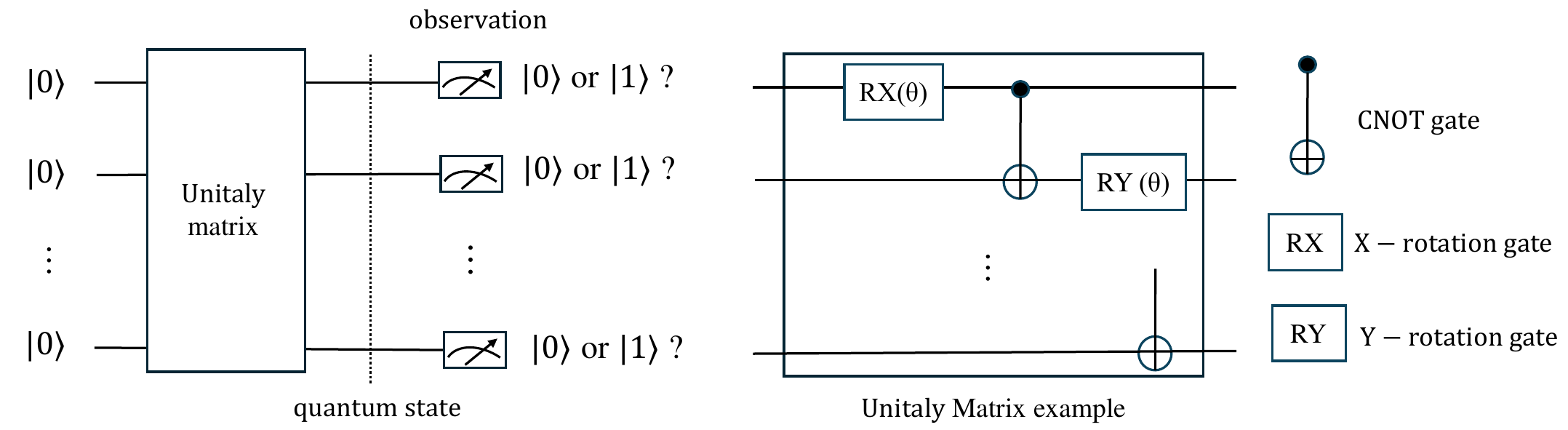}
    \caption{Gate structure for encoding and weight matrix}
    \label{supfig:quantum_circuit}
\end{figure}

We explain angle encoding to a quantum circuit from features.
Here, $\vx=(x_1,\dots,x_d)^\top\in[-1,1]^n$ is input vector, qubit number is $n$.
Since a single-qubit state is $\ket{\psi} = \cos\frac{\theta}{2}\ket{0} + e^{j\phi}\sin\frac{\theta}{2}\ket{1}$ via Bloch sphere parameters.
We preserve the normalization of the amplitudes from the input data as a function of the angles $\theta$ and $\phi$.
This naturally introduces inverse trigonometric functions; the initial angle for the $i$-th qubit is
\begin{align*}
\theta_i = \arcsin(x_i),
\qquad
\phi_i = \arccos(x_i^2).
\end{align*}
We apply the unitary matrix $U_i(\vx)$ to the $i$-th qubit as shown in~\fref{supfig:quantum_circuit}.
\[
  U_i(\vx) \;=\; R_Z^{(i)}(\phi_i)\,R_Y^{(i)}(\theta_i)
\]
In the implementation, operations act on the state from the right, so $R_Y$ is applied before $R_Z$; in matrix notation, the order is written as $R_Z R_Y$.
As the rotations are independent, the overall encoding circuit $U_{\mathrm{enc}}(\vx)$ is
\[
  U_{\mathrm{enc}}(\vx) \;=\; \bigotimes_{i=1}^{n} U_i(\vx)
  \;=\; \bigotimes_{i=1}^{n} \Bigl[\,R_Z^{(i)}\!\bigl(\phi_i\bigr)\,R_Y^{(i)}\!\bigl(\theta_i\bigr)\Bigr].
\]
The mapping from the basis state $\ket{0}$ is given by
\[
  R_Y(\theta_i)\ket{0}=\cos\frac{\theta_i}{2}\ket{0}+\sin\frac{\theta_i}{2}\ket{1},
  \
  R_Z(\phi_i)\ket{0}=e^{-i\phi_i/2}\ket{0}, R_Z(\phi_i)\ket{1}=e^{+i\phi_i/2}\ket{1},
\]
Therefore,
\[
  U_i(\vx)\ket{0}
  = e^{-i\phi_i/2}\cos\frac{\theta_i}{2}\,\ket{0}
      + e^{+i\phi_i/2}\sin\frac{\theta_i}{2}\,\ket{1},
  \,\because\
  \theta_i=\arcsin(x_{i}),\phi_i=\arccos(x_{i}^{2}).
\]

The trainable unitary matrix $W(\boldsymbol{\uptheta})$ is appended after the feature encoding $U_i(\vx)\ket{0}$. The circuit $W(\boldsymbol{\uptheta})$ is capable of adapting through training, regardless of the initial state $\ket{0}$.
\[
  U(\vx,\boldsymbol{\uptheta}) \;=\; W(\boldsymbol{\uptheta})\,U_{\mathrm{enc}}(\vx).
\]

The resulting parameterized quantum state $\ket{\psi(\vx,\boldsymbol{\uptheta})}$ is

\[
  \ket{\psi(\vx,\boldsymbol{\uptheta})}
  \;=\;
  U(\vx,\boldsymbol{\uptheta})\,\ket{0}^{\otimes n}
  \;=\;
  W(\boldsymbol{\uptheta})\,
  \Bigl(\bigotimes_{i=1}^{n} R_Z^{(i)}\!\bigl(\arccos(x_{i}^{2})\bigr)\,R_Y^{(i)}\!\bigl(\arcsin(x_{i})\bigr)\Bigr)\,
  \ket{0}^{\otimes n}.
\]
\subsubsection{Gradients per bag on Q-LLP}
\begin{figure}[t]
    \includegraphics[width=\linewidth]{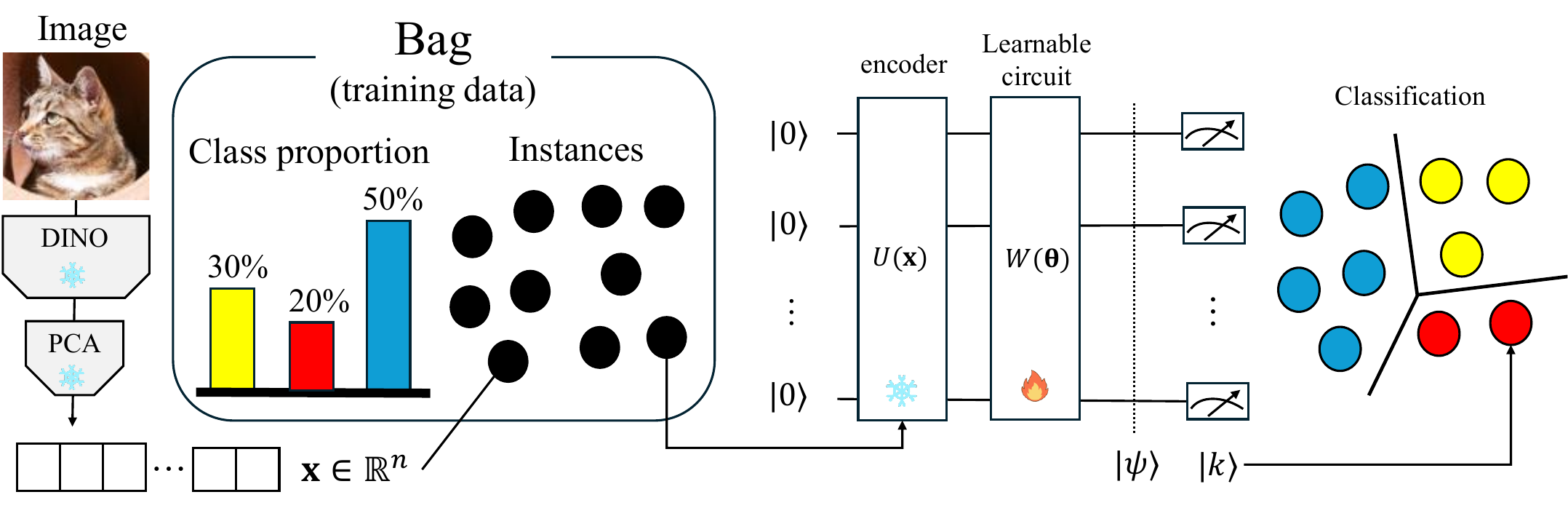}
    \caption{Overview of our Q-LLP. }
    \label{fig:method_qllp}
\end{figure}

We explain how to compute the parameter gradients in a weakly supervised setting, which is learning from the label proportion (LLP)~\cite{LLP-2008-origin}.
LLP allows the strength of the supervision signal to be adjusted.
As shown in Fig.~\ref{fig:method_qllp}, images are first processed by feature extraction, which are subsequently reduced to the qubit dimension $n$ using Principal Component Analysis (PCA), resulting in an instance representation $\vx \in \mathbb{R}^n$.
The multiple input instances in a bag $\mathbb{X}=\{\vx_1, ...,\vx_i,...,\vx_m\}$ are encoded from the initial state $\ket{0}$ using $R_Y$ and $R_Z$ gates via angle encoding, resulting in $U_{enc}(\vx)\ket{0}^{\otimes n}$.

Each instance goes through the weight unitary matrix $W(\boldsymbol{\uptheta})$, which consists of multiple layers of $R_X$, $R_Z$ with learnable parameters $\boldsymbol{\uptheta}=[\theta_1,...] \ (\theta \in \mathbb{R})$, and dense matrix gates for quantum superposition between each qubit. Then it obtains a quantum state as follows.

\[
\ket{\psi(\boldsymbol{\uptheta})}=W(\boldsymbol{\uptheta})U(\vx)\ket{0}^{\otimes n}
\]

From the states $\{\ket{\psi_1},...,\ket{\psi_i},...,\ket{\psi_m}\}$ of the bag, we construct the density matrix

\[
\hat{\rho}(\boldsymbol{\uptheta}) = \frac{1}{m} \sum_{i=1}^{m} \ket{\psi_i(\boldsymbol{\uptheta})} \bra{\psi_i(\boldsymbol{\uptheta})}
\]

From the true $C$-class proportion $\vp=(p_1,\dots,p_q)$ of the bag, the target density matrix is

\[
\rho := \sum_{k=1}^C p_k \ket{k}\bra{k}
\]

Here $\ket{k}$ is the computational basis state corresponding to class $k$.
From the viewpoint of reducing the number of qubits and computation cost, the class $k \in {1,\dots,C}$ is represented in binary with $n=\lceil\log_2 C\rceil$ qubits, and the basis state corresponding to that bitstring is associated with the class. For example, in the case of qubit number $n=2$ and class number $C=4$,
$\ket{00}\to\mathrm{class}\ 1,
\ket{01}\to\mathrm{class}\ 2,
\ket{10}\to\mathrm{class}\ 3,
\ket{11}\to\mathrm{class}\ 4.$

\label{sec.bag_grad_loss}

We use the Hilbert-Schmidt inner product for fidelity $F(\boldsymbol{\uptheta})$ between the target density matrix $\rho$ and the density matrix $\hat{\rho}(\boldsymbol{\uptheta})$:

\begin{align*}
F(\boldsymbol{\uptheta})
&= \mathrm{Tr}\bigl(\rho \hat{\rho}(\boldsymbol{\uptheta})\bigr)\\
&= \mathrm{Tr}\!\left(
  \rho \,\frac{1}{m}\sum_{i=1}^m \ket{\psi_i(\boldsymbol{\uptheta})}\bra{\psi_i(\boldsymbol{\uptheta})}
  \right) \\
&= \frac{1}{m}\sum_{i=1}^m
   \mathrm{Tr}\!\left(\rho \,\ket{\psi_i(\boldsymbol{\uptheta})}\bra{\psi_i(\boldsymbol{\uptheta})}\right).
\end{align*}

By applying the properties of the trace \(\mathrm{Tr}(A \, \ket{\psi}\bra{\psi}) = \bra{\psi}A\ket{\psi}\),

\begin{align*}
F(\boldsymbol{\uptheta})
&= \frac{1}{m}\sum_{i=1}^m \bra{\psi_i(\boldsymbol{\uptheta})}\,\rho\,\ket{\psi_i(\boldsymbol{\uptheta})}
\end{align*}

The loss function per bag is defined as

\[
\mathcal{L}_{\text{bag}}(\boldsymbol{\uptheta}) = -\log \bigl(F(\boldsymbol{\uptheta})+\varepsilon\bigr)
\]

where $\varepsilon$ is a small constant for numerical stability. The training goal is to minimize loss function: $\text{min} (\mathcal{L}_{\text{bag}}(\boldsymbol{\uptheta}))$.
The gradient of each parameter $\theta_j$ is obtained as $\frac{\partial \mathcal{L}(\boldsymbol{\uptheta})}{\partial \theta_j}$.

\[
\frac{\partial \mathcal{L}_{\text{bag}}(\boldsymbol{\uptheta})}{\partial \theta_j}
= -\frac{1}{F(\boldsymbol{\uptheta})+\varepsilon}
\,\frac{\partial F(\boldsymbol{\uptheta})}{\partial \theta_j}.
\]

The derivative of fidelity is

\[
\frac{\partial F(\boldsymbol{\uptheta})}{\partial \theta_j}
    =\frac{\partial}{\partial \theta_j}\frac{1}{m}\sum_{i=1}^m \bra{\psi_i(\boldsymbol{\uptheta})}\,\rho\,\ket{\psi_i(\boldsymbol{\uptheta})}
    =\frac{1}{m}\sum_{i=1}^m \frac{\partial}{\partial \theta_j}\bra{\psi_i(\boldsymbol{\uptheta})}\,\rho\,\ket{\psi_i(\boldsymbol{\uptheta})}
\]

From the derivative of a product$(f \cdot g)' = f' \cdot g + f \cdot g'$

\[
\frac{\partial}{\partial \theta}\bra{\psi}\,\rho\,\ket{\psi}\
    =\bra{\frac{\partial}{\partial \theta}\psi}\,\rho\,\ket{\psi} + \bra{\psi}\,\rho\,\ket{\frac{\partial}{\partial \theta}\psi}
\]

From $\rho^{\dagger} = \rho \Rightarrow \overline{\rho} = \rho \Rightarrow \bra{a}\,\rho\,\ket{b} = \overline{\bra{b}\,\rho^{\dagger}\,\ket{a}} = \overline{\bra{b}\,\rho\,\ket{a}}$

\begin{align*}
\frac{\partial}{\partial \theta}\bra{\psi}\,\rho\,\ket{\psi}
    &= \bra{\tfrac{\partial}{\partial \theta}\psi}\,\rho\,\ket{\psi}
      + \overline{\bra{\tfrac{\partial}{\partial \theta}\psi}\,\rho\,\ket{\psi}} \\
    &= 2 \,\mathrm{Re}\bra{\tfrac{\partial}{\partial \theta}\psi}\,\rho\,\ket{\psi}
    &\hfill \because \theta \in \mathbb{R}
\end{align*}

Therefore,

\[
\frac{\partial F(\boldsymbol{\uptheta})}{\partial \theta_j}
    = \frac{2}{m}\sum_{i=1}^m \mathrm{Re}\bra{\frac{\partial}{\partial \theta_j}\psi(\boldsymbol{\uptheta})}\,\rho\,\ket{\psi(\boldsymbol{\uptheta})}
\]

By defining $g_j=\bra{\frac{\partial}{\partial \theta_j}\psi(\boldsymbol{\uptheta})}\,\rho\,\ket{\psi(\boldsymbol{\uptheta})}$ as the gradient of the i-th instance $\vx_i$, we obtains the final gradient $\vg_i = [g_1, \dots ]$ for $\boldsymbol{\uptheta}=[\theta_1,...]$.
Bag size $m=1$ means supervised learning. Note, however, that an increase in bag size is not monotonically related to an increase in classification difficulty.

\subsection{Parameter efficiency and KTA}

\begin{figure}[t]
    \includegraphics[width=\linewidth]{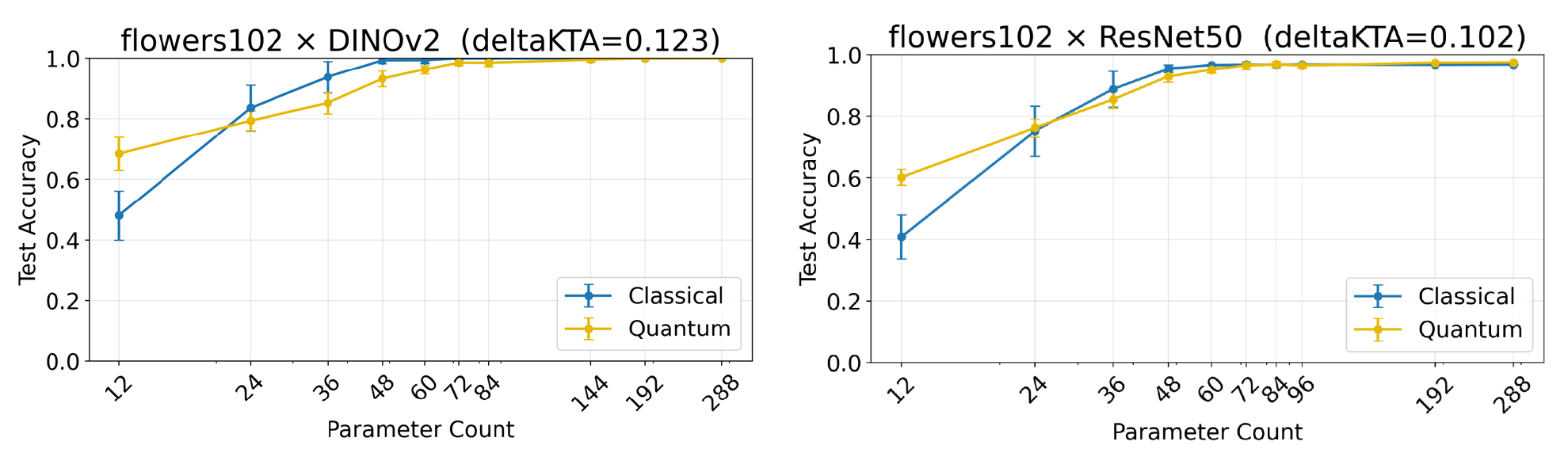}
    \caption{
    Test accuracy as a function of parameter count for classical and quantum classifiers when the quantum feature map does not improve kernel–label alignment ($\Delta$KTA $< 0$).
    In these settings, the classical representation already provides a favorable feature geometry, and both models exhibit similar scaling behavior with respect to parameter count.
    }
    \label{fig:vsParam_2}
\vspace{-3mm}
\end{figure}

We select two representative cases, Flower102$\times$DINOv2 and Flower102$\times$CLIP, and find that only a small subset shows strong positive alignment across the 70 evaluated combinations, and analyze how the parameter count affects the classification performance of classical and quantum classifiers.
Figures~\ref{fig:vsParam_2} show the test accuracy as a function of the number of trainable parameters for two dataset–feature extractor pairs.

The subplot reports the difference in Kernel Target Alignment between quantum and classical feature representations,
\[
\Delta \mathrm{KTA}
=
\mathrm{KTA}_{\mathrm{quantum}}
-
\mathrm{KTA}_{\mathrm{classical}} .
\]

Datasets with positive $\Delta\mathrm{KTA}$ indicate that the quantum feature map improves the alignment between the kernel matrix and labels, suggesting that the induced Hilbert space provides a more separable representation.

Among the datasets with large $\Delta$KTA, only \textit{Flowers102} with CLIP features shows a clear advantage of the quantum classifier at small parameter counts.
This behavior suggests that these features are particularly compatible with the quantum feature space.

In contrast, the STL10 results exhibit large variance for the classical model.
Across 30 random seeds, the optimization often remains near chance-level accuracy, indicating strong sensitivity to initialization.

Overall, these results suggest that the benefit of quantum classifiers depends on the compatibility between the input feature geometry and the quantum feature space.
When the representation aligns well with the quantum feature map, competitive performance can be achieved with fewer parameters.

\subsection{Sensitivity to bag size}
\label{sec:appendix_bag_size}

\begin{figure}[t]
    \centering
    \includegraphics[width=\linewidth]{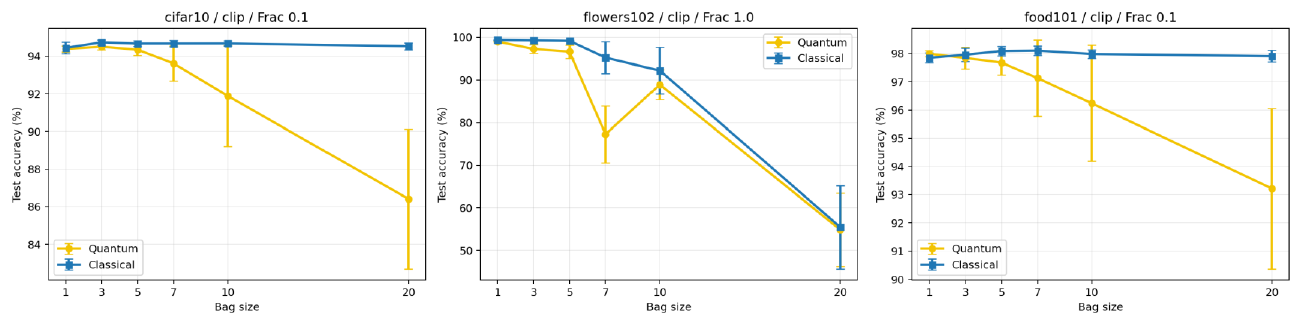}
    \caption{
    Test accuracy (%
    We report mean and standard deviation over random seeds.
    The setting is fixed to: CIFAR10$\times$CLIP (Frac $=0.1$), Flowers102$\times$CLIP (Frac $=1.0$), and Food101$\times$CLIP (Frac $=0.1$).
    }
    \label{fig:sup_bag_size}
\end{figure}

Figure~\ref{fig:sup_bag_size} shows how the bag size $m$ affects the final test accuracy in LLP training.
Overall, the classical model is relatively stable across $m$ in CIFAR10 and Food101, while the quantum model tends to degrade as $m$ increases, with larger variance at large bag sizes.

For CIFAR10 and Food101, the quantum model is competitive at small bag sizes ($m\in\{1,3,5\}$), but accuracy decreases for larger bags, suggesting that aggregating many instances into one supervision unit can weaken optimization signals for the current quantum parameterization.
For Flowers102, both models drop at very large bag size ($m=20$), indicating that excessive bag aggregation can harm optimization regardless of model family in this setting.

These results are consistent with the gradient formulation in Sec.~\ref{sec.bag_grad_loss}: increasing $m$ averages instance-level contributions within each bag, which can reduce gradient magnitude and increase optimization difficulty when instance diversity inside bags is high.
Practically, this suggests using small-to-moderate bag sizes as a robust default, and tuning $m$ jointly with feature extractor and dataset characteristics.

\subsection{Computation resouces}
All experiments were run on a Linux server equipped with 8 NVIDIA A100-SXM4 GPUs (40 GB memory each), two AMD EPYC 7742 64-core CPUs, and 1.0 TiB system memory. Feature extraction used CUDA, where available, with a batch size of 128 and 4 data-loading workers. The quantum classifier was executed using Qulacs on a CPU.

Training and inference for a quantum classifier take about 24.5 hours in a setting with 1 seed and 500 iterations.

\subsection{Licenses}
We provide a summary of the subset and the license for the dataset used in our experiments.

\begin{table}[htb]
\centering
\scriptsize
\begin{tabular}{p{2.5cm}p{2.9cm}p{4.8cm}}
\toprule
Dataset/resource & Version/subset used & License status \\
\midrule
CIFAR-10 & 4-class subsets & Non-Commercial Use \\
Fashion-MNIST & 4-class subsets & MIT \\
SVHN & 4-class subsets & Non-Commercial Use \\
STL-10 & 4-class subsets & Non-Commercial Use; images acquired from ImageNet \\
Oxford-IIIT Pet & 4-class subsets & CC BY-SA 4.0 \\
Oxford Flowers 102 & 4-class subsets & Non-Commercial Use \\
Food-101 & 4-class subsets & Scientific fair use \\
MedMNIST & multiple subsets & CC BY 4.0; DermaMNIST is CC BY-NC 4.0 \\
QMNIST & 4-class subsets & BSD-style license \\
\bottomrule
\end{tabular}
\caption{Dataset license status summary.}
\end{table}

\end{appendices}

\end{document}